# Biosensors and Machine Learning for Enhanced Detection, Stratification, and Classification of Cells: A Review

Hassan Raji[a], Muhammad Tayyab[a], Jianye Sui[a], Seyed Reza Mahmoodi[a], Mehdi Javanmard[a*]

[a]*Department of Electrical and Computer Engineering, Rutgers University, Piscataway, NJ, 08854, USA*

**Abstract**

Biological cells, by definition, are the basic units which contain the fundamental molecules of life of which all living things are composed. Understanding how they function and differentiating cells from one another therefore is of paramount importance for disease diagnostics as well as therapeutics. Sensors focusing on the detection and stratification of cells have gained popularity as technological advancements have allowed for the miniaturization of various components inching us closer to Point-of-Care (POC) solutions with each passing day. Furthermore, Machine Learning has allowed for enhancement in analytical capabilities of these various biosensing modalities, especially the challenging task of classification of cells into various categories using a data-driven approach rather than physics-driven. In this review, we provide an account of how Machine Learning has been applied explicitly to sensors that detect and classify cells. We also provide a comparison of how different sensing modalities and algorithms affect the classifier accuracy and the dataset size required.

*Keywords:* Biosensors; Machine Learning (ML); Neural Networks; Deep learning; Microfluidics; Support Vector Machine;

## 1. Introduction

There are still numerous healthcare challenges despite many advances in technology.[1] Biosensors as analytical tools have several applications in different fields such as diagnostics and disease monitoring which play a significant role in this respect. [2] A biosensor is a system that can do selective and quantitative detection of an analyte or biomarker utilizing a biorecognition part and a signal transduction part.[3] These analytes can be non-biological such as drugs, toxins, dissolved gases, etc. or biological such as cells, proteins, DNAs, etc.[4] In a biosensor, once an analyte of interest is detected by a biorecognition element, the presence of the analyte is confirmed by a transducer quantitatively or semi-quantitatively. Then, the generated signal due to the recognition event is converted to an output signal. There has been a significant interest in biosensors in analytical systems for medical diagnostics.[5] Also, biosensor measurements in microchannels, considering the small volume of the fluid and its relative large surface area, has also been quite popular and provides a number of benefits including a high rate of heat and mass transfer.[6] Biosensors that detect cells have garnered special interest in the last few decades with the advent of technologies which automate and miniaturize its different components. In particular, they are important for the diagnosis and detection of various diseases which include but are not limited to: Sickle cell Disease[7], Acute Myeloid Leukemia[8], and metastatic cancers[9] through detection of Circulating Tumor Cells (CTCs). An important stage when using these sensors in a clinical setting is converting the data obtained from these biosensors into useful information by classifying the cells into different categories. For example, Circulating Tumor Cells need to be identified and separated from Red Blood Cells. There are a number of qualities which make a biosensor that detects cells more popular including rapid performance and response[10], high specificity[11], high sensitivity[12]. Also, other beneficial qualities include continuous measurement of analyte without involving experienced personnel[13], range[14], response time[15], stability[16], low cost[17], and accuracy.[18] Processing of the generated data from biosensors can be considered as an important stage that effectively influences the improvement of the above-mentioned qualities.

---

[*] Corresponding author. Tel.: +1-(848) 445-3382;
*E-mail address:* mehdij@alumni.stanford.edu.

Machine learning, a subset of artificial intelligence, is a framework allowing algorithms to learn automatically from data. Many techniques based on machine learning (ML) have been shown to solve significantly difficult tasks in the real world. They are especially applicable to tasks that require learning a variety of patterns obtained from data. The reason such models can work extremely efficiently and with human-level performance is the fact that while using such methods, the given problem is defined in a precise mathematical framework. That framework uses large amounts of either labeled or unlabeled data, and then some general probabilistic algorithm is applied to find patterns in the dataset. Evidently, this can have numerous advantages as well as several drawbacks. These advantages include the fact that in many applications since some general model is used, there is no further need for hand-engineered expert knowledge which can be quite expensive or even ambiguous. For medical applications in particular, it has been shown that such methods can not only significantly outperform human-engineered expert knowledge, but they are also able to discover new knowledge.[19] Another advantage is that sometimes these methods discover patterns that could not have been discovered independently and might have seemed irrelevant at first. This overall makes them much more scalable compared to human intervened knowledge discovery. However, these approaches have some drawbacks as well. For example, in many applications, they are very heavy computations that takes several weeks for some models to train. More importantly, they require costly predefined labels in some supervised scenarios. Additionally, some applications are extremely sensitive to the choice of architectures or the hyperparameters are chosen. These drawbacks are being actively improved. As an example, in many classical classification problems, quite simple methods such as logistic regression or Support Vector Machine (SVM) have been shown to perform extremely well. For more complex tasks, more complex neural net-based architectures can be required.

In some biosensors, a large amount of data is generated quickly at the output, and the analysis of this data requires further processing by an experienced user that can lead to errors. Processing by a person can take time to analyze data, which can greatly reduce the efficiency of the biosensor. On the other hand, ML can identify features and trends, and can also provide understandable output. A quick web search shows that the application of Machine Learning in biosensors have seen an exponential rise in the last decade.

Other review papers have reviewed deep learning applied on microfluidics and image cytometry, but no paper specifically discussed the application of the broader concept of ML on biosensors detecting cells using various sensing modalities.[20],[21] In this paper, a review of ML publications on biosensors detecting cells is discussed whilst some pieces of useful information will be provided for biosensor engineers and scientists who want to use ML in their research. In this regard, an overview on main ML concepts is firstly discussed. The papers in this review are divided into four main categories based upon the detection mechanism used. These include: Electrical Detection, Optical Flow Cytometry, Microscopy-based detection, and Smartphone-based detection.

## 2. Overview of main concepts in machine learning

### *2.1. Supervised ML*

In the supervised approach, a set of pre-defined labels should also be fed into the algorithm alongside the input dataset. It is then the algorithm's job to first extract meaningful features from raw data and then find the best parameters that are able to predict the mentioned labels on a separate test dataset as accurately as possible.

### *2.2. Unsupervised ML*

Unlike supervised ML approaches, in unsupervised approaches there are no pre-defined labels available in advance. As a result, algorithms themselves should discover the meaningful representations of the data that are useful per se or in some other downstream task, for instance a classification or regression problem as before. These algorithms can take different approaches towards finding such meaningful representations such as by probabilistic density estimation, clustering or latent variable modeling, etc. For instance, E.M (Expectation Maximization) is one of such methods.

### *2.3. Artificial Neural Networks*

One branch in machine learning which has recently gotten significant attention is called Artificial Neural Nets (ANN). These methods are loosely inspired by the inner functioning of the human brain, but in fact such methods apply many highly nonlinear and complex functions, a.k.a neurons, to the input data in a parallel fashion. These complex nonlinear dynamics allow them to extract much more complex and useful feature representations from raw data, thus leading to more useful representations and significantly better performances in many complex tasks. The process of learning within such ANN models is in fact the finding of optimal parameters for synaptic weights of the neurons in order to gain a reasonable accuracy. Also, it is necessary to mention that in most ANN architectures, more than one layer of neural operations are cascaded to make them solve more complex tasks, thus giving them the name "Deep Learning models".

### *2.4. Convolutional Neural Networks*

A specific form of ANNs is called Convolutional Neural Nets (CNNs). These architectures are specifically designed for image-based tasks such as image/video classification, object detection, tracking, recognition, etc. although they have been applied to other problems as well. In contrast to Feed Forward Neural nets, these architectures use specially designed cells that utilize a convolution operation. More specifically, the learnable weights of the network are the parameters of a set of convolutional kernels which are convolved with the input images or the outputs of each layer. Such architectures were initially designed for problems focusing on images, mainly because they take advantage of the effect of spatial-invariance in the images as well as the importance of locally-neighboring features. A result, they convolve the same shared parameters across the whole image.

### *2.5. Support vector machine*

Support vector machine (SVM) is one of the most commonly used methods for many supervised classification tasks where a set of n-dimensional data-points are given as the input, each accompanying a true label. The goal of SVM is to find hyper-planes (generalization of lines and planes to higher dimensional spaces) that can accurately separate these data-points. For instance, in 2D space, this hyper-plane would be a simple line which divides the space into two sub-regions, each corresponding to a different class. These hyperplanes are defined by a set of points called the support vectors.

## 3. Machine Learning in Different Biosensing Techniques

### *3.1. Microscopy-based Detection*

Microscopy-based detection implies the use of a microscope to obtain images or videos of cells. These images or videos need to be processed to identify and quantify cells. ML has tremendous power in the analysis of microscopic image data by making accurate predictions on large sample datasets. ML algorithms eliminate much of the manual steps required to process data, thereby reducing the processing time and eliminating human error. In the next following paragraphs, we present sensing approaches based on ML algorithms on data obtained using this detection method.

Neural networks are often utilized in image analysis, and thus, this technique also draw a lot of attention to the studies of cells using microscopic detection. Koohababni et al. utilized Mixture Density Networks (MDNs) to identify cell nucleus.[22] MDNs are suitable candidates for mapping single inputs to multi outputs. So, in their work, these NNs were used to detect several seeds in an image path. Features were learned by a CNN in which images were used as the input datasets. Furthermore, MDN detected nucleus within the image patch by a Gaussian distribution. Their presented method was able to identify cell nucleus in colorectal histology images. In another study, the authors used neural networks (NN) in inline holography microscopy for high-speed cell sorting.[23] They showed that this label-free imaging technique can be applied for ultrafast, cell sorting with a high accuracy. A CNN-based single-frame super-resolution processing proposed by Huang et al. for lensless microfluidic cell counting with a demonstrated lensless blood cell counting protype.[24] The CNN-based single-frame SR processing improved the low-resolution cell images with lower hardware cast thus making it potentially a viable candidate for use in point-of-care diagnostics. Mayerich and colleagues presented a method for cell soma detection in Knife-Edge Scanning Microscopy (KESM) using ML.[25] The high throughput of this data required cell classification to be performed at a high rate. In this paper, they used pattern recognition employing a multi-layer feed-forward neural network to accurately locate neuron positions in the rat brain. They demonstrated that the accuracy of their algorithm exceeded the performance of standard feature detection algorithms and can be implemented on commonly available and affordable hardware architectures. Other researchers demonstrated the use of a NN to simulate the movement and behaviour of red blood cells in blood plasma.[26] In this study, the NN was taking a numerical simulation as an input. Alternatively, the input could also be a video recorded from an actual biological experiment. Their results indicated that for uncomplicated box channels, there was no advantage of using this method instead of fluid streamlines. However, in a more complicated geometry, the NN performance showed a significant improvement. Such a simulation could be used for optimizing the microfluidic channel geometry. A combination of feature selection algorithm and NN classifiers was carried out in another research. The objective of this study was to recognize five types of white blood cells in the peripheral blood.[27] For this purpose, nucleus and cytoplasm were segmented using the Gram-Schmidt method and snake algorithm. Moreover, three kinds of features (morphological, textural, and color) were extracted from the segmented areas. Next, the best features were selected using Principal Component Analysis (PCA). Finally, five types of white blood cells were classified using Learning Vector Quantization neural network (LVQNN). Falk et al. proposed plugin in a software package for cell detection and cell segmentation based on deep learning allowing users to employ this plugin without having knowledge of ML.[28] Unlike previous similar software packages, their plugin, U-Net, had the capability to be trained and adapted to new sets of data and tasks by ImageJ® software interface.

Deep learning as a powerful tool to overcome image segmentation was also applied in the study performed by Van Valen and colleagues. The authors demonstrated that deep CNN is able to successfully segment and classify different mammalian cells. This deep learning technique was accurate, curated segmentation results in a short period of time, segmented variety of cell types, and differentiated different types of cell lines from each other.[29] Zhang and colleagues have recently demonstrated a novel cell detection and cytometry technique by incorporating magnetically modulated lensless imaging.[30] A deep learning-based classifier was empoyed to enhance the specificity of their cytometer which also allowed to detect MCF-7 circulating tumor cells based on their spatio-temporal features under a controlled magnetic force. In another study, Akram et al. presented a CNN-based method providing cell segmentation proposals. These proposals initially represented bounding boxes utilizing a fully CNN (FCN) and then predicted segmentation masks for bounding boxes using another CNN.[31] They compared their proposed techniques with other conventional cell detection and segmentation methods and concluded that their method has a better performance in terms of common evaluation parameters. Similarly, Xie et al. used a deep learning-based object detection method, Faster Region-based CNN, along with a transfer learning process to detect cells in microscopic images.[32] By conducting analysis on 314 images, 50 for training and 314 for testing, they reported a miss rate of 1.3% and a detection accuracy of 98.4%. In another biosensor study, Faster Region-based CNN was applied for cell detection by segmentation and classification, to cell detection.[33] Their experiments showed that cells can be detected in microscopic images using Faster R-CNN. Furthermore, this technique improved cell detection performance, saved time, and was easily implemented. Another research group, developed a high-throughput and automated RBC classification method utilizing patient-specific microscopy images.[34] In this work, initially a hierarchical RBC patch extraction method was used for the sickle cell disease (SCD) sensing. Additionally, a shape-invariant RBC patch normalization technique was employed for the input of deep nets enabling to save time during learning and training procedures and to exclude unnecessary background patches.

In a biosensor research, T-cells and B-cells were distinguished in a pillar-based microfluidic cell counting system by applying a SVM classifier based on the histogram of oriented gradients (HOG) and color distribution features (Fig. 1).[35] First, a linear-kernel SVM was trained to detect cells from a background in dual dyed images. Subsequently, the cells in a single dye image were identified by the first SVM based on HOG features found in the image using a sliding window method. At last, a Radial Basis Function (RBF)-kernel SVM was trained with the color information of found cells to differentiate T-cells from B-cells.

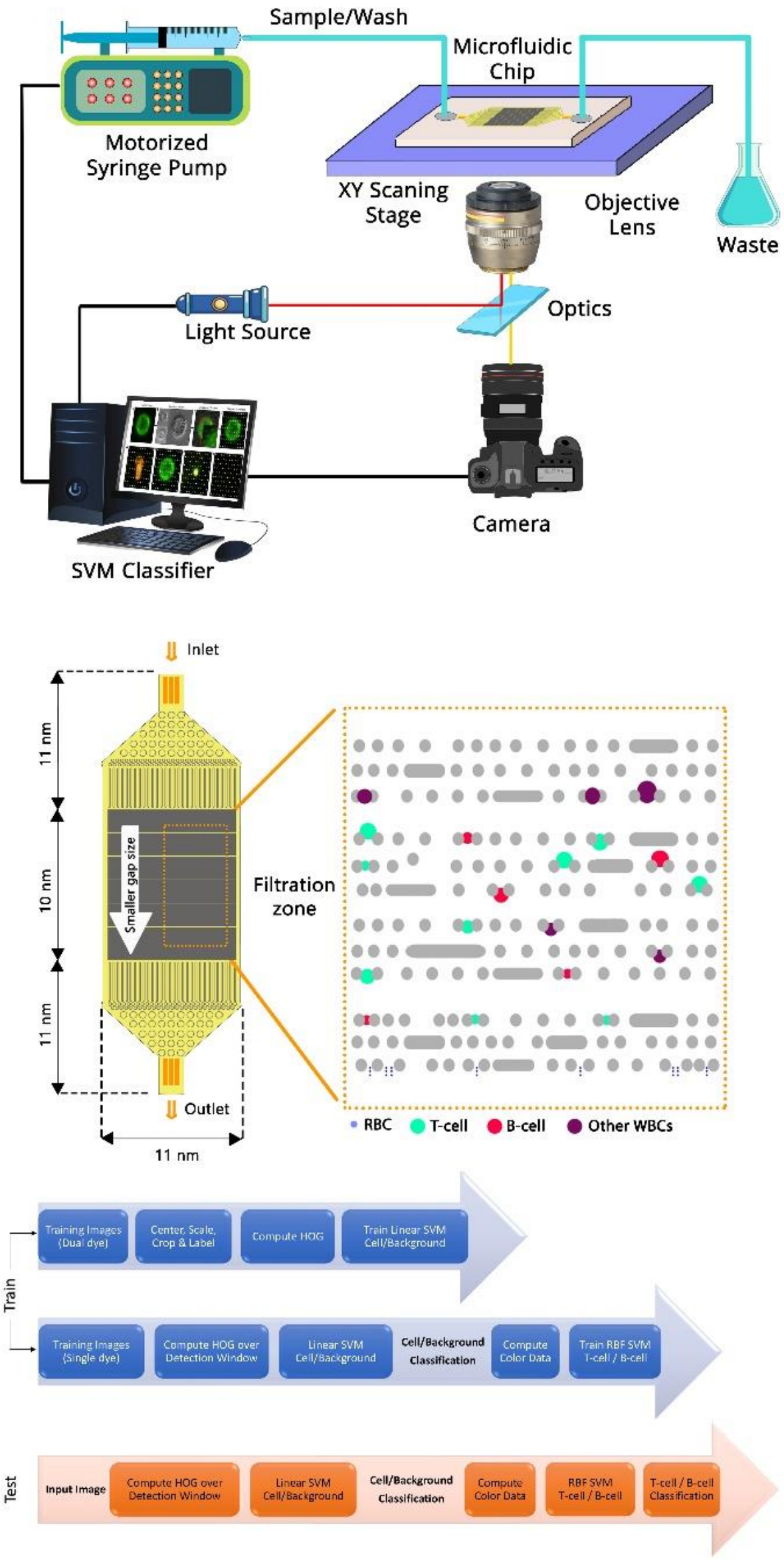

**Fig. 1** Outline of the Microscopy-based system proposed in reference [35]. Blood was injected into the device through inlet while leukocytes were trapped in different zones based on the deformability and size difference. The proposed experiment setup and block diagram of cell detection using ML is also shown in this figure. In the block diagram of the cell detection framework, it can be inferred that using Support Vector Machine (SVM), training images are centered, cropped, and labeled. The Histogram of Gradients (HOG) and color data was computed using the processed images for the classification of cells. Adapted from [35].

SVM can be used for image classification, and also for detecting and differentiating the cells in post-processed images. A support vector machine was employed by Long et al. with an iterative training procedure to detect unstained viable cells in bright field images.[36] They proposed a novel algorithm called “Compensatory Iterative Sample Selection" to handle the extremely unbalanced training sample set, which made the decision boundary more accurate. Uslu et al. developed computer vision-based algorithms to quantify the leukemia cells captured and separated by immunomagnetic beads.[37] SVM was implemented to quantify target cells in the images captured by a bright field microscope. To rebalance the dataset, where non-cell images were the majority, non-cell images were down sampled before the training, radial basis function was selected as the kernel function. In another work performed by Guo et al. an SVM was applied to analyze the intensity and phase images acquired by the optofluidic time-stretch quantitative phase microscopy.[38] They proposed a high-throughput label-free single-cell method for screening lipid-producing microalgal cells by optofluidic time-stretch quantitative phase microscopy. Features extracted from the images were used in the classification of nitrogen-sufficient and nitrogen-deficient E. gracilis cells. In another study, a computerized detection of Acute lymphoblastic leukemia using microscopic images was investigated.[39] K-means algorithm was performed after image processing to segment cell nuclei. The geometric features and statistical features were extracted for classification. Means of SVM classifier were used to classify cancerous and noncancerous cells. The cells were further classified into subtypes by a multi-SVM classifier. The accuracies of the two classifiers were both above 95%. Similarly, ML has also been used in image-based screening of bacterial growth. A live microscopy detection system of bacterial growth was presented by Wand et al.[40] They showed that this imaging platform was able to analyze the time-lapsed holographic images using deep NNs for rapid detection of bacterial colonies within 7–10 h. Guo and colleagues proposed a high-throughput label-free single-cell method for screening lipid-producing microalgal cells by optofluidic time-stretch quantitative phase microscopy.[41] A SVM was applied to analyze the intensity and phase images acquired by the optofluidic time-stretch quantitative phase microscopy. 188 features extracted from the images were used in the classification of nitrogen-sufficient and nitrogen-deficient E. gracilis cells. It achieved an 2.15% error rate in cell classification.

Some other ML algorithms are also employed in analyzing the data obtained by microscopic image cytometry. Huang et al. demonstrated the use of a technique based on Extreme Learning Machines (ELM) for single-frame super-resolution processing applied on a microfluidic contact imaging cytometer platform.[42] Compared with the commercial flow cytometer, less than 8% error was observed for the absolute number of microbeads. They demonstrated in another paper that by mixed flowing of HepG2 and Huh7 cells as the inputs, the developed scheme achieved 23% better recognition accuracy compared to the one without error recovery. Whereas, it also achieved an average of 98.5% resource-saving compared to the previous multi-frame super-resolution processing.[43] Autoencoders are considered an unsupervised learning technique since they don't need explicit labels to train on. However, to be more precise they are self-supervised because they generate their own labels from the training data. A microfluidics-based platform for single-cell imaging in-flow and subsequent image analysis using Variational Autoencoders (VAE) for unsupervised characterization of cellular mixtures was demonstrated In Constantinou’s paper.[44] Heterogeneous mixtures of yeast species were classified with 88% accuracy. Microfluidic Imaging Flow Cytometry (MIFC) is an emerging method of microscopic imaging, which aims to reduce the complexity of the tasks involved in cytometry by combining flow cytometry with digital microscopy.[45] This technique promised significantly higher throughput and was easy to set up with minimal expenses in Kalmady et. al study.[45] This group employed MIFC for obtaining images instead of image cytometry. They proposed a transfer learning and ensemble learning-based approach for the automation of cytopathological analysis of Leukemia cell-line images. Compared to earlier works, the use of fine-tuned features from a modified deep NN for transfer learning provided a substantial improvement in performance. In another example, Tang Yu et al. employed image processing algorithms to classify yeast cells in a microfluidic channel.[46] They compared linear support vector machine (LSVM), distance-based classification (GED), and k-nearest-neighbor (KNN) classifiers. It was also shown that these three classifiers had similar accuracy in their biosensor among which KNN being the most versatile classifier, and SVM has the fastest processing time.

A new segmentation algorithm for the classification of five types of white blood cells by Su et and colleagues.[47] Their segmentation algorithm was based on finding a discriminating region of white blood cell tones in the color space. In their study, three different NN-based classifiers of MLP, SVM and HRCNN were adopted for classifying white blood cells. It was shown that the proposed system incorporated with a trained MLP can reach the highest performance. In another label-free approach, Go et al. used digital in-line holographic microscopy (DIHM) paired with ML models to identify and classify different types of erythrocytes: discocytes, echinocytes, and spherocytes.[48] Four different models were used to determine the best algorithm: Support Vector Machine (SVM), Decision Trees, Linear Discrimination Classification (LDC), and k-nearest neighbor (KNN) classification. The decision trees exhibited the best identification performance for the training sets (n=440,98.18%) and test sets (n=140, 97.37%). In terms of two ML-based approaches, namely ELMSR and CNNSR, a research was conducted to solve the low-resolution problem in a lensless microfluidic imaging using CMOS image sensors for blood cell counting.[49] In this paper, low-resolution lensless of cell image was the input and an improved high-resolution cell image was the output. At the end, cell resolution was improved 400% while the cell counting results were in line with commercial flow cytometers. The same group proposed a single frame lensless microfluidic imaging where a ML algorithm, ELM-SR, was used for recovering high-frequency details existing in the low resolution frames.[50]

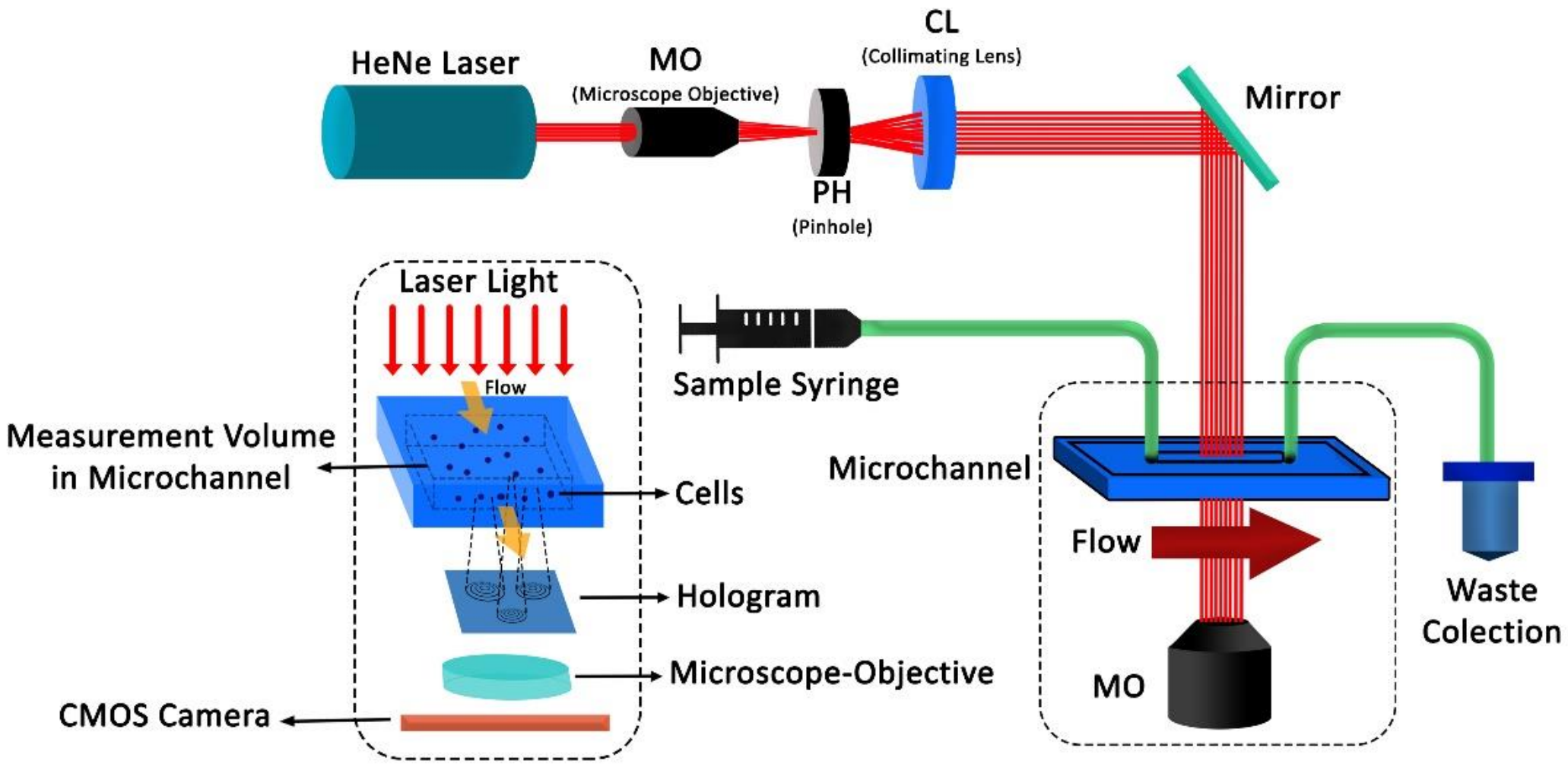

**Fig. 2** Inline digital holography microscopy (DHM) utilized in [51] by Singh and colleagues [51] for characterizing cells in flow. As shown, in this figure, experimental arrangement of inline-DHM is shown which enabled recording holograms of cells in bulk flow along with multiple experimental parameters. The output data was used in a classifier enabling detection of tumor cells. Adapted from [51].

Singh et al. employed ML-based gating criteria to differentiate tumor MCF-7 circulating tumor cells from blood cells when flowing through a microchannel (Fig. 2).[51] First, classifiers were developed using extracted features from training sets of blood cells and tumor cell lines. This classification system was tested with CTS spiked in a background of blood cells and was able to identify tumor cells concentration of 10 cells per ml, with false positive rate of 0.001%. Mao et al. proposed an microscopic image-based circulating tumor cells (CTC) detection employing an SVM classifier with hard-coded Histograms of Oriented Gradients features and a CNN classifier with automatically learned features.[52] Their classifiers applied to a challenging dataset which showed that it detected CTC automatically in a minimally invasive way. This image-based CTC detection was independent of the cell marker expression and was not limited to any cancer type.

### *3.2. Optical Flow Cytometry*

A microfluidic flow cytometer is an integrated system which consists of microchannels for flow and optical sensors for detection. Typically, the cell is detected using scattered light from laser beams illuminating the cells flowing through the detection chamber in a microchannel. Ideally, the biosensor would be portable, easy to operate, and suitable for use as a point-of-care diagnostic device.

Various research groups applied deep learning with their microfluidic flow cytometers to analyze the single-cell images for cells classification .[53] A CNN classifier allowed them to identify the class of 21-by-21 pixel single-cell images in less than 1 millisecond. This classifier utilized image extraction and recognition by training the network on acquired image datasets of cells. With the help of classifier, the cytometer accurately counted and identified label-free flowing cells by using a live video stream of a large-volume sample. Soldati et al. classified droplet content by combining multiple ML algorithms.[54] By integrating computer vision techniques, automatic classification of droplets was carried out using CNNs with an accuracy of 96%. In addition, this group utilized NN for object detection and were able to segment the images of droplets and cells in order to measure their relative volumes. It corrected estimation of ECAR up to 20%. A deep learning pipeline was proposed by Li and his colleagues. It operated directly on the measured signal from the time-series waveforms of an imaging flow cytometer. The features were extracted employing the model itself (Fig. 3).[55] In this study, a long feature extraction and signal processing steps were entirely avoided so that the time of cell analysis was reduced significantly. It has been shown that this method was able to classify OT-II WBCs and SW-480 epithelial cancer cells with a high accuracy of 95%. Hence, cell sorting was performed in orders of magnitude faster than previous studies, enabling real-time label-free cell sorting.

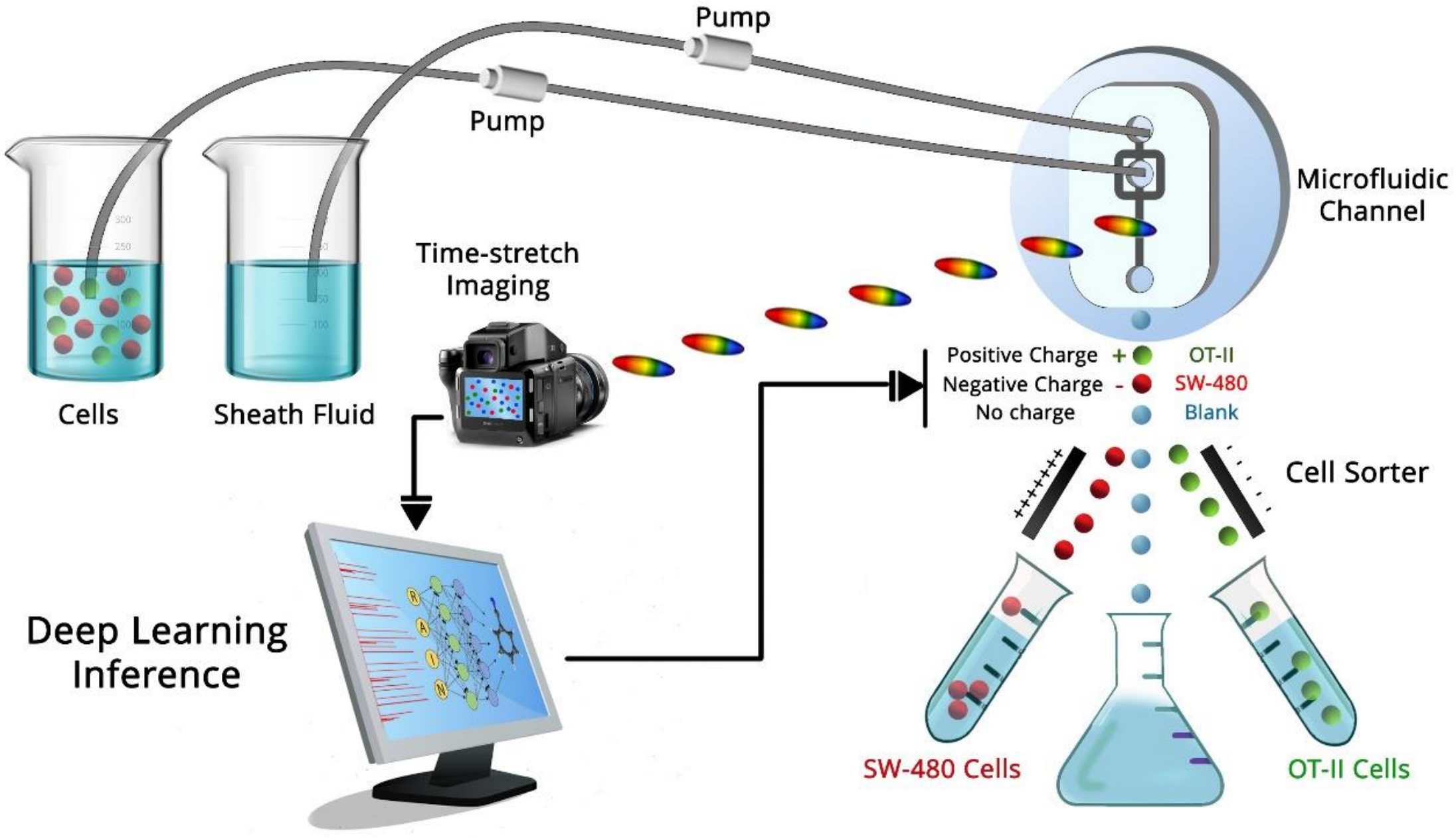

**Fig. 3** Overview of the application of deep learning in flow cytometry presented in [55]. In this research, hydrodynamic focusing mechanism was employed in a microfluidic channel to align the cells in the centerline of the main channel. Line images of the cells in the channel was captured by time-stretch imaging system by the rainbow pulses. These rainbows were images of cells which were flowing into the microchannel at high speed. Consequently, without further signal processing, the waveforms of time-stretch imaging were outputted to a deep NN where cell classification was carried out rapidly with high accuracy. Different cells were charged and then categorized before being separated into different collection tubes. Adapted from [55].

In a microfluidic-based imaging flow cytometry (IFC) technique, an accurate classification framework was presented for the first time. It was based on deep learning for unstained IFC data for three unstained, unlabeled, unstained leukemia cell lines.[56] They demonstrated that instead of using conventional fine segmentation and explicit feature extraction, by using deep-learning algorithms, coarsely localized cell lines can be successfully classified. Sun et al. used a deep CNN to learn the biological characteristics of 2D light scattering patterns in the azimuthal and polar angle from a microfluidic cytometer and ultimately identified label-free lymphocytic leukemia cells. Their deep learning network accurately detected Jurkat and BALL-1 cells with an accuracy of 0.932, and the sensitivity and specificity were 0.92 and 0.94453. An ANN was used in Glushkova and colleagues' work to count blood cells based on the light signal, when cells passed through a microchannel.[57] This system was used for classification of leukocytes, erythrocytes and platelets of blood samples. Another research group, presented a label-free technique that used a digital inline holographic microscopy for cell imaging, integrating NN for high-speed classification.[58] Their method classified cells at a much shorter process delay in comparison with previous studies due to the use of immediate holographic interface pattern as the the input of NN. An 89% accuracy was obtained from network simulations for a ternary classification task of WBCs. Chen et al. integrated feature extraction and deep learning with high-throughput quantitative imaging enabled by photonic time stretch, achieving high classification accuracy (95.5%).[59] Their system captures quantitative optical phase and intensity and extracts multiple biophysical features of individual cells. These biophysical measurements thus form a hyperdimensional feature space in which supervised ML is performed for cell classification.

Among the studies employing ML for optical flow cytometers, SVM is one of the most popular algorithms. In an study, SVM algorithm was applied to analyze MFC dataset in order to detect minimal residue disease in acute myeloid leukemia and myelodysplastic syndrome patients automatically.[60] The original raw data were encoded using a multivariate Gaussian mixture model and then fed into the SVM classifier. They validated this with a large-scale clinical data and clinical outcome. Another research group developed an in vivo Photoacoustic Flow Cytometry (PAFC) system to achieve in vivo melanoma inspection.[61] They implemented a support vector machine algorithm to discriminate signals and noises based on the continuity, amplitude, and photoacoustic waveform pulse width extracted from photoacoustic waves. A model accuracy of 92% was accomplished. Lin et al. developed a label-free light-sheet microfluidic cytometer for single cell analysis by two-dimensional (2D) light scattering measurements.[62] Incorporating the cytometer with SVM algorithms, a high accuracy was achieved in automatic classification of senescent and normal human fibroblasts. Four parameters (contrast, correlation, energy and homogeneity) were calculated for light scattering patterns and used as features in the SVM classifier. A linear kernel function was adopted with 5-fold cross validation. The SVM was used by Toedling and colleagues for automatic detection of leukemic cells from patients' bone marrow and peripheral blood samples in flow cytometry readouts.[63] Manually gated leukemic cells were recovered by SVM with 98.87% specificity and 99.78% sensitivity which showed the potential of a well-established multivariate-analysis technique.

### *3.3. Smartphone-based Detection*

Smartphone-based sensors are closely related to the microscopy-based sensors since they replace the microscope with the smartphone cameras, which are often supplemented by an attachment and have the same output i.e. image. They are becoming increasingly popular because of their small footprint and widespread availability of smartphones. Furthermore, they eliminate the need for specialized optical equipment like microscopes and spectroscopes by substituting it with relatively inexpensive and portable attachments. Smartphone-based biosensors utilizing ML, therefore, have tremendous promise for being used as point-of-care diagnostic devices with minimal training and knowledge required for operation.

A cost-effective method proposed by De Haan et al. was capable of automatic screening of sickle cells (SC) in a deep learning framework.[64] The framework included two complementary deep NN (Fig. 4). The first one standardized and enhanced blood smear images from a smartphone microscope while the second one acted on the output of the first image and performed the semantic segmentation between SC and healthy cells in a blood smear. Furthermore, the segmented images were utilized for the diagnosis of SC disease and achieved an accuracy of 98%.

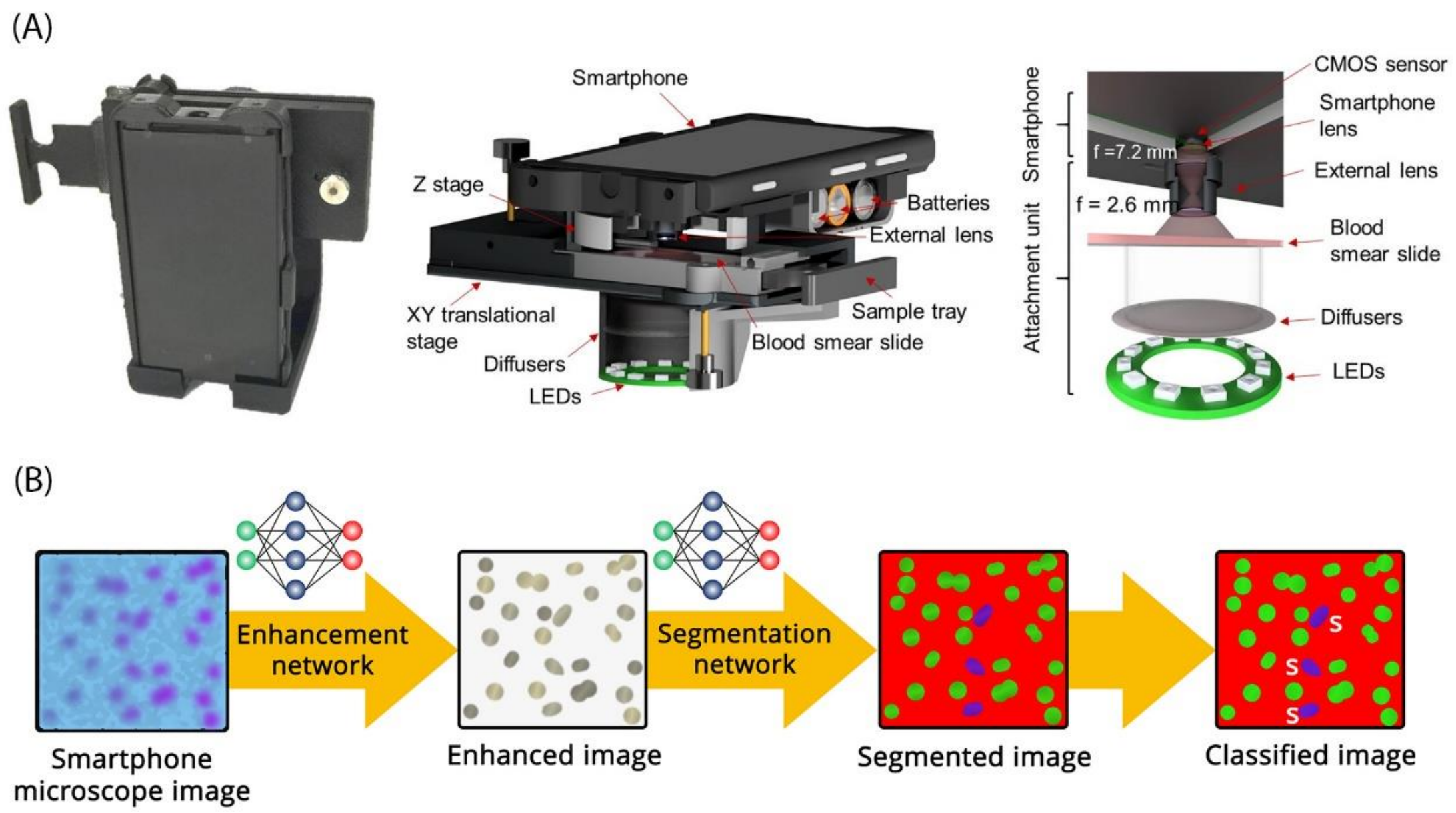

**Fig. 4** Overview of smartphone-based biosensor employed by de Haan and colleagues. [64] (A) Photograph of the smartphone-based system, the overall design, and the light path is shown from left to right. Reprinted from [64] (B) workflow of deep learning process is presents. This learning algorithm has been used sickle cell analysis to enhance blood smear images and carry out semantic segmentation between SC and healthy cells. Adapted from [64].

A comparison of different ML algorithms was carried out for waterborne pathogen (Giardi) detection using a smartphone- based setup.[65] The accuracy and the Area Under the ROC curve (AUROC) of different ML models were compared including, but not limited to SVM, nearest neighbors and ensemble methods. All the models had a classification accuracy above 81%, while the AUROC values were greater than 0.7. The best predictive performance was obtained using bagged trees (Ntree=400). Fine and cubic kNN classifiers provided fast fitting speeds, but their predictive accuracy was relatively poor. On the other hand, SVM and bagged ensemble classifiers were promising at their prediction accuracy, while their training speeds were slower.

### *3.4. Electrical Detection*

Electrical detection refers to the use of electrical circuits to obtain data in the form of electrical signals. These signals can be impedance, voltage, current or any other electrical signal. Impedance is generally the most commonly used parameter to identify and quantify cells. When a cell passes through the electrodes in a microfluidic channel, a change in impedance occurs. The output signal is determined by the cell's properties such as cell size, conductivity, and permittivity. Electrical Detection of cells has many advantages over traditional optical detection. Since there is no need for bulky optical equipment, electrical detection devices usually have a small footprint and are less expensive. In the following paragraphs, we present biosensors utilizing ML techniques for electrical detection of various biomarkers.

Integrated with ML algorithms, a microfluidic impedance cytometry for real-time, label-free multiparametric characterization of biological cells.[66] In this study, a recurrent NN was designed to predict cell diameter, velocity, and position from electric current signals, measured by a microfluidic impedance chip. The trained network was able to characterize geometric and electrical properties of beads, red blood cells, and yeasts with a good accuracy and a unitary prediction time of 0.4 ms. Zhao et al. designed a new microfluidic impedance cytometry with crossing constriction microchannels, which allows quantifying the cellular electrical markers.[67] Using an equivalent circuit model, they translated the measured impedance values to specific membrane capacitance and cytoplasm conductivity. A NN-based pattern recognition was used to classify tumor cell lines and tumor cells with epithelial-mesenchymal transitions. Precise measurement of mechanical and/or electrical properties of cells or cell components yields useful information on the physiological and pathological state of cells and is critical for cell classification. Yang et al. extracted deformability, electrical impedance and relaxation index of single cells from impedance spectroscopy measurements with self-aligned 3D electrodes.[68] They demonstrated the ability of their system to detect and classify cells using a back propagation NN completely based on the biophysical properties of the cells. In another study, a microfluidic constriction channel was designed to measure single-cell electrical properties.[69] A back propagation NN was used for cell classification based on three parameters of diameter, specific membrane capacitance, and cytoplasm conductivity. Finally, they showed that cell classification success rate significantly improved when information additional to cell size was included.

In Chen and his colleagues' work, osteoblasts and osteocytes were classified using a two-layer back propagation NN70. The input data had three groups of parameters measured on cells, namely, transit time, impedance amplitude ratio, and cell elongation length. Their results suggested that biomechanical and bioelectrical parameters, when used in combination, provided a higher cell classification success rate than using alone. In another study, a microfluidic system was presented for cell type classification based on size-independent electrical properties, specific membrane capacitance and cytoplasm conductivity.[70] Two lung tumor cell lines were classified using a two-layer back propagation NN. The NN-based classification resulted in a fairly acceptable classification success rate of 65.4% (CSpecific Membrane), 71.4% (σcytoplasm), and 74.4% (CSpecific Membrane combined with σcytoplasm). A microfluidic system proposed by Zhang et al. with a constriction channel. The channel was marginally smaller than the RBC's diameters which was used to classify adult and neonatal RBCs using a back propagation NN through their biophysical properties (mechanical and electrical).[71] Electrical measurements were performed to characterize these properties. The input data had three group of parameters (transit time, amplitude ratio and phase increase). The results showed that when these parameters were used in combination, yielded a relatively higher classification accuracy (84.8%) than the time each parameter was used alone. Recently a study published where the authors used Quadratic Discriminant Analysis (QDA). This is a type of supervised ML algorithm that helped them extract six features from Red Blood Cells (RBCs) and yeast cells using Impedance micro-cytometry. They achieved the maximum test accuracy (99%) by using four features on RBCs. They also demonstrated the efficacy of their platform by classifying different cancer subtypes. The accuracy decreased when more than four features was used. It was because of overfitting of the model to the training data.[72]

A study conducted cancer drug efficacy analysis using multifrequency impedance cytometry, measuring the impedance of a single cell at several discrete frequencies.[73] Support vector machine algorithm was implemented to help differentiate alive cells from dead cells. Song et al. employed a support vector machine algorithm to help identify differentiation states of stem cells based on impedance signals collected by the microfluidic electrical impedance flow cytometer at 50 kHz, 250 kHz, 500 kHz and 1 MHz.[74] Another research group discriminated strains of E. coli K-12, E. coli O157: H7, and Salmonella Thompson using a multichannel immunosensor incorporated with multiclass support vector machines.[75] Gini-SVM framework was adopted to design multiclass SVMs. To evaluate the performance, a 100-fold cross-validation procedure was implemented.

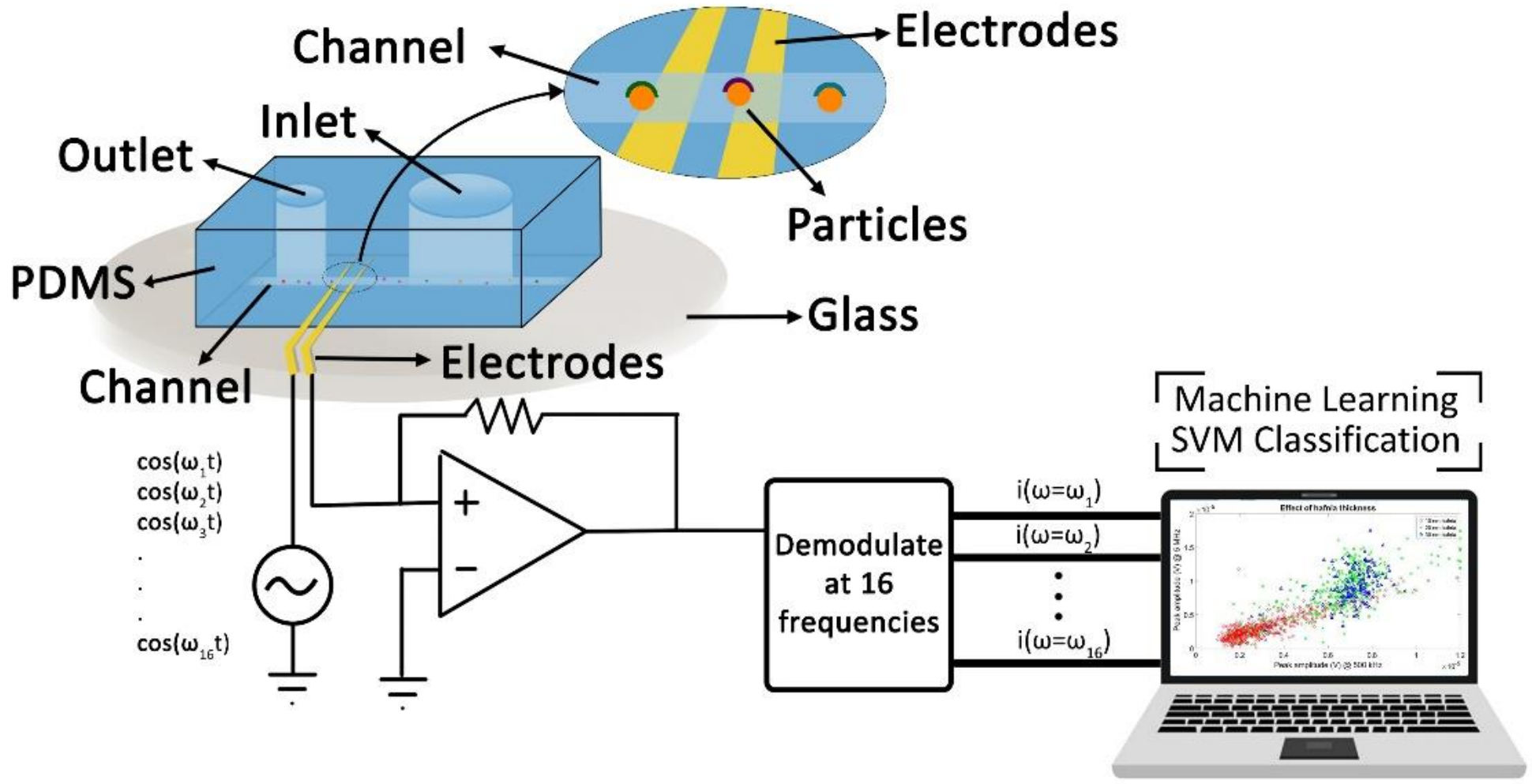

**Fig. 5** Schematic diagram of an electrical impedance cytometer. As cells flow from the inlet to the outlet in these biosensors, the change in impedance is measured by a lock-in amplifier. The lock-in amplifier can apply signal in different frequencies at a time. The data are then recorded and analyzed using SVM. Adapted from [76].

Detection and enumeration of circulating tumor cells from red blood cells were performed in research using a micropore-based microfluidic impedance cytometer.[38] The peak amplitude and the pulse bandwidth of signal pulses were analyzed by SVM to differentiate cancer cells from red blood cells. Radial basis function (RBF) was appointed as the kernel function. The results of the proposed microfluidic sensor combined with SVM showed a good agreement with the results of a commercial flow cytometer. Wang and colleagues proposed a sensitive multiplex self-referencing SERS pathogen detection scheme.[77] A linear kernel-based SVM in conjunction of PCA was performed for rapid discrimination and classification of target bacteria with a detection accuracy above 95%. An approach for hematocrit estimation from the transduced anodic current curves introduced in a study. The curves were obtained by glucose-oxidase reaction in the strip-type electrochemical biosensors.[78] The support vector machine was implemented for regression with the target value of accurate hematocrit values measured by a hospital analysis system.

**TABLE 1 Comparison of ML efficacy of different Biosensors for Cellular Analysis**

| Author & Year of Publication | Training Algorithm | Detection Technique | Sample Type | Cell type | Approximate Classifier Accuracy (Maximum)* | Dataset Size | Ratio |
|---|---|---|---|---|---|---|---|
| Dataset size = number of cells | | | | | | | |
| Toedling [2006][63] | Multivariate Classification, SVM | Optical Flow Cytometry | Yeast Culture | Yeast Cells | 82% | 120 Cells | 6.83E-01 |
| Tabrizi [2010][27] | Neural Networks | Microscopy-based | Biological Tissue | Neuron Cell Soma | 92.80% | 2158 Cells | 4.30E-02 |
| Chen [2011][79] | Neural Networks | Electrical | Cell co-culture | RBC and HepG2 Tumor Cells | 99% | 11909 Cells | 8.31E-03 |
| Yu [2011][46] | GED, SVM and KNN | Microscopy-based | Blood | Red Blood Cells | 84.80% | 166326 Cells | 5.07E-04 |
| Mayerich [2011][25] | Neural Networks | Microscopy-based | Stem Cell Solution | Mouse Embryonic Carcinoma Cells (P 19) | 95% | 98 Signals (Cells and Beads) | 9.69E-01 |
| Van Valen [2011][29] | Deep learning | Microscopy-based | Cell Culture | CRL-5803 cells and CCL-185 Cells | 74.40% | 976 Cells | 7.62E-02 |
| Zheng [2012][71] | Neural Networks | Electrical | Cell Suspension | ML-2 and HL-60 Cells | 93% | 6647 Cells | 1.40E-02 |
| Song [2013][74] | SVM | Electrical | PBS | HepG2 and RBC Cells | 92.00% | 3698 Cells | 2.49E-02 |
| Zhao [2013][70] | Neural Networks | Electrical | Blood and Bone Marrow Smears | Acute lymphoblastic leukemia Cells | 97% | 958 Cells | 1.01E-01 |
| Zheng [2013][69] | Neural Networks | Electrical | Blood | Monocytes, Granulocytes, and Lymphocytes | 89% | 7500 Cells | 1.19E-02 |
| Huang [2014][42] | Neural Networks (ELM-SR) | Microscopy-based | Blood | Leukemia Cells | 99.46% | 10000 Cells | 9.94E-03 |
| Moradi [2015][39] | SVM | Microscopy-based | Cell Culture | Fluo-N2DL-HeLa, PhC-HeLa and Hist-BM Cells | 96.90% | 34060 Cells | 2.84E-03 |
| Schneider [2015][23] | Neural Networks | Microscopy-based | Lysis of RBCs | MDA-MB-231 and MCF7 Cells | False Positive Rate of at Most 0.001% | 100000 Cell Training Sets | N/A |
| Ni [2016][80] | SVM | Optical Flow Cytometry | Cell lines | Leukemia cell lines(K562, MOLT, and HL60) | 97.60% | 618 Cells | 1.58E-01 |
| Akram [2016][31] | Deep Learning | Microscopy-based | Cell lines | leukemia cell lines HL60, MOLT, and K562 | 99.52% | 618 Leukemia cells | 1.61E-01 |
| Singh [2017][51] | Decision Tree | Microscopy-based | Buffer | Lung cancer cell lines ofH1299 and A549 | 90.90% | 100,000 Cells | 9.09E-04 |
| Gopakumar [2017][56] | Deep Learning | Optical flow cytometry | Online Dataset | Colorectal Adenocarcinoma Cells | Precision=0.788 | 29756 Nuclei | N/A |
| Kalmady [2017][45] | Neural Networks | Optical Flow Cytometry | Blood | Erythrocytes and Platelets | Minimum Error=0.003 | > 40k Cells | N/A |
| Zhao [2018][67] | Neural Networks | Electrical | Buffer | Normal Human Fibroblasts (NHFs) and Senescent Human Fibroblasts (SHFs) | 88% | 480 Cells (240 Normal Human Fibroblasts (NHFs) and 240 Senescent Human Fibroblasts) | 1.83E-01 |
| Alemi [2018][22] | Deep Learning | Microscopy-based | Buffer | T47D Cancer Cells | 95.90% | >1000 Cells | 9.59E-02 |
| Glushkova [2018][57] | Neural Networks | Optical Flow Cytometry | Biological Tissue | Circulating Tumor Cells (CTCs) | 85% | 600000 Cells | 1.42E-04 |
| Lin [2018][62] | SVM | Optical Flow Cytometry | Blood | RBCs and Yeast Cells | RMSE=1.2 um for particle size | 17000 Single Particle Signals (Beads and Cells) | N/A |
| Ahuja [2019][73] | SVM | Electrical | Yeast Culture | Yeast Cells | 82% | 120 Cells | 6.83E-01 |
| Fu [2019][61] | SVM | Optical Flow Cytometry | Biological Tissue | Neuron Cell Soma | 92.80% | 2158 Cells | 4.30E-02 |
| Honrado [2020][66] | Neural Networks | Electrical | Cell co-culture | RBC and HepG2 Tumor Cells | 99% | 11909 Cells | 8.31E-03 |
| **Dataset size = Number of Images** | | | | | | | |
| Long [2006][36] | SVM | Microscopy-based | Cell Culture | B-cell Lymphoma Cells | 94% | 59 Images | 1.59 |
| Su [2014][47] | Neural Networks | Microscopy-based | Blood | White Blood Cells | 99.11% | 450 Images | 2.20E-01 |
| Mao [2015][52] | SVM, CNN | Microscopy-based | Blood | CTCs (Breast Cancer) | 91.20% | 45 Images | 2.03 |
| Heo [2017][53] | CNN | Optical Flow Cytometry | Microparticles in Buffer | K562 and RBCs | 93.30% | 6000 Images | 1.55E-02 |
| Xu [2017][34] | CNN | Microscopy-based | Blood | Sickle Cells (RBCs) | 89.28% | > 7000 single RBC images | 1.28E-02 |
| Go [2018][48] | Decision Tree, SVM, kNN, Linear Discriminant Classification | Microscopy-based | Blood | Erythrocytes: Discocytes, Echinocytes, Spherocytes | 97.37% | 630 Images | 1.55E-01 |
| Turan [2018][35] | SVM | Microscopy-based | Blood | T and B Cells | 94% | 420 Scan Images | 2.24E-01 |
| Soldati [2018][54] | CNN | Optical Flow Cytometry | Blood | CTCs and CD45 Cells | 90.20% | 500 Images | 1.80E-01 |
| Xia [2019][32] | Deep Learning | Microscopy-based | Buffer | White Blood Cells | 98.40% | 364 Images | 2.70E-01 |
| Sun [2019][81] | Deep Learning | Optical flow cytometry | Buffer | T Cells and B Cells (Acute Luekemia) | 93.20% | 2400 Images | 3.88E-02 |
| Uslu [2019][37] | SVM | Microscopy-based | Buffer | B Lymphoblast | 87.40% | 100000 Sub-images | 0.000874 |
| **Dataset size = Other** | | | | | | | |
| Park [2008][78] | SVM, ELS-ELM, RLS-ELM, ELM | Electrical | Blood | Red Blood Cells | RMSE=0.74 | 199 Blood Samples | N/A |
| Ko [2018][60] | SVM | Optical Flow Cytometry | Blood | Acute Myeloid Leukemia (AML) and Myelodysplastic Syndrome (MDS) Cells | 92.40% | 5333 MFC Data Points | N/A |
| Li [2019][55] | Deep learning | Optical Flow Cytometry | Buffer | OT-II White Blood Cells and SW-480 Epithelial Cancer Cells | 95.74% | 6700 Data Points | N/A |
| de Haan [2019][64] | Deep Learning | Smartphone-based | Whole blood | Sickle Cells and Healthy RBCs | 98% | 96 Patients' Blood Smear | N/A |
| Zhang [2019][30] | Deep Learning | Microscopy-based | Whole Blood | MCF-7 Cancer Cells | 78% | 17,447 Videos | N/A |

## 4. Conclusions

ML has become a useful tool in analyzing and classifying the data obtained from biosensors for cellular analysis. Based on the papers we discussed here, we see that both SVM and ANN are the prevalent techniques which are effective in automating the classification of various cell types except for microscopy-based biosensors. This may be due to the fact that these are the most widely known techniques for researchers working in fields related to biosensors. In the case of microscopy-based biosensors, various NN architectures are preferred over SVM and other methods for classification of different cell types.

Table I compares important characteristics for several papers that are discussed in this review. It is divided into 2 main categories for easily comparing each method's efficacy. In the first part, we list the papers in which the number of cells are specified as the dataset size. The second part of the table contains the papers in which the number of images is considered as the dataset size. Other than these 2 main categories, we have listed a few papers in which the dataset neither corresponds to the number of images nor the number of cells. The last column of Table I is a ratio of the classifier accuracy (in percentage) to the dataset size. A higher number, therefore, indicates that the accuracy achieved corresponds to a relatively small dataset.

We also noted that Deep Learning and ANN have grown more popular recently with the majority of the newer publications using these methods. Another interesting observation is that biosensors utilizing electrical detection methods rarely employ deep learning as the analytical tool for classifying cells. This may be due to the fact that deep learning is data hungry and databases for electrical biosensing data are not yet established. On the other hand, there is a plethora of datasets easily available that may be used as training samples for image and optical detection of various biomarkers.

To summarize, the use of ML algorithms in biosensors have huge benefits that automate the cumbersome and complicated process of extracting, processing and analyzing data that is generated by the biosensors. Such an automation eliminates the need for an experienced professional to make sense of the data and moves us closer to providing Point-of-care health solutions in environments that have low resources. Although ML algorithms have been around for a while now and have huge benefits, the techniques discussed here mostly utilize code and require certain Integrated Development Environments (IDE's) e.g. Python, MATLAB etc. for their use. Researchers should consider packaging of the softwares into a GUI which will make these relatively simple to interact with and less formidable.

## Conflicts of interest

Disclosure of potential conflict of interest: M. Javanmard and J. Sui has a pending patent for "Use of Multi-Frequency Impedance Cytometry in Conjunction with Machine Learning for Classification of Biological Particles" and M. Javanmard has equity in Rizlab Health Inc., a company dedicated to commercialization of a microfluidic hematology analyzer.

## Acknowledgements

This research was sponsored by National Science Foundation Awards 1556253 (IDBR), 1846740 (CAREER), and 1711165 (ECCS-CCSS) and also the Defense Advanced Research Projects Agency (DARPA) Biological Technologies Office (BTO) Electrical Prescriptions (ElectRx) program managed by Dr. E. V. Giesen through the DARPA Contracts Management Office Grant/Contract Number HR0011-16-2-0026 and Contract Number N660011824018. The views, opinions and/or findings expressed are those of the authors and should not be interpreted as representing the official views or policies of the National Science Foundation, the Department of Defense, or the U.S. Government.